\documentclass[a4paper]{jpconf}
\usepackage{graphicx}

\begin{document}
\title{Universal quantum gates for Quantum Computation on magnetic systems ruled by Heisenberg-Ising interactions}

\author{Francisco Delgado}
\address{Departamento de F\'isica y Matem\'aticas, Escuela de Ingenier\'ia y Ciencias, Tecnol\'ogico de Monterrey, Campus Estado de M\'exico, Atizap\'an, Estado de M\'exico, CP. 52926, M\'exico.}
\ead{fdelgado@itesm.mx}

\begin{abstract}
The gate version of quantum computation exploits several quantum key resources as superposition and entanglement to reach an outstanding performance. In the way, this theory was constructed adopting certain supposed processes imitating classical computer gates. As for optical as well as magnetic systems, those gates are obtained as quantum evolutions. Despite, in certain cases they are attained as an asymptotic series of evolution effects. The current work exploits the direct sum of the evolution operator on a non-local basis for the driven bipartite Heisenberg-Ising model to construct a set of equivalent universal gates as straight evolutions for this interaction. The prescriptions to get these gates are reported as well as a general procedure to evaluate their performance. 
\end{abstract}

\section{Introduction}\label{sec1}
Quantum computation and quantum information are modern developments taking advantage from the Quantum Mechanics features to propose technological applications. In their quantum gate version, the gate construction on specific physical systems is a central aspect to develop those applications. The requirement of entanglement as resource implies the introduction of physical interactions between the parts of the system. This requirement introduces in the dynamics other natural basis than computational basis, which is only natural for single and isolated qubits. For this reason, in general, it is not easy fit the evolution into the theoretical gate constructions inspired on the classical computational elements \cite{deutsch1,steane1}. The same is true for more complex gates involving multiqubit systems. Thus, several attempts to define general ways to construct gates are in order. These approaches uses unitary factorization, asymptotic approximations, etc. \cite{schende1,li1,delgadoC, paige1}. Still the most common approach is the construction of universal sets of gates whose composition lets to generate any other gate.

In this work, we propose and develop a set of universal gates easily constructed for bipartite magnetic systems ruled by the anysotropic Heisenberg-Ising interaction including driven magnetic fields in the $x, y, z$ directions ($h=1,2,3$):

\begin{eqnarray} \label{hamiltonian}
H_h&=&\sum_{k=1}^3 J_k {\sigma_{1_k}} {\sigma_{2_k}}-{B_{1_h}} {\sigma_{1_h}}-{B_{2_h}} {\sigma_{2_h}} 
\end{eqnarray}

\noindent this model comprises several models reported in the literature \cite{berman1,wang2,dales1,dales2,kamta1,sun1,zhou1}. Their $SU(4)$ dynamics exhibits a block form when it is expressed on the non-local basis of Bell states \cite{delgadoA}. As it has been proved, the general blocks in that decomposition has the form $U(2)=U(1) \times SU(2)$:

\begin{eqnarray}\label{sector}
{s_{h_{j}}} &=& {e^{i {\Delta_h}_\alpha^+} \left(
\begin{array}{cc}
{{e_h}_\alpha^\beta}^* & -q i^h {d_h}_\alpha   \\
q {i^*}^h {d_h}_\alpha & {{e_h}_\alpha^\beta}    
\end{array}
\right) } = {e^{i {\Delta_h}_\alpha^+}} \left( \cos {{\Delta_h}_\alpha^-} {{\bf I}_{h_j}} - i \sin {{\Delta_h}_\alpha^-} {\bf n} \cdot { {{\bf S}_{h_j}}} \right) 
\end{eqnarray}

\noindent where ${\bf n}=(q {{b_h}_{-\alpha}} \sin \frac{h \pi}{2}, q {{b_h}_{-\alpha}} \cos \frac{h \pi}{2}, \beta {{j_h}_{-\alpha}}), \alpha=(-1)^{h+j+1}, \beta=(-1)^{j(h+l_j-k_j+1)}, q=\beta (-1)^{h+1}$. $h$ is the direction of magnetic field and $j=1, 2$ a position label for each block in the whole evolution matrix. $k_j, l_j$ are the labels for its rows. There, ${{\bf I}_{h_j}}$ and ${{{\bf S}_{h_j}}}$ are a set of extended Pauli matrices stating a basis for each $SU(2)$ block $j$ applied on definite pairs of selectable Bell states as function of $h$. This basis could be understood too as noise or error basis for deviations from the evolution prescriptions (such as the traditional $X, Y$ and $Z$ are noise effects in the computational basis, dephasing and flipping noise). The parameters:

\begin{eqnarray} \label{defs}
{e_h}_\alpha^\beta = \cos {\Delta_h}_\alpha^- + i \beta {j_{h_{-\alpha}}} \sin {\Delta_h}_\alpha^- \label{delta}, \quad {d_{h_\alpha}} = {b_{h_{-\alpha}}} \sin {\Delta_h}_\alpha^-  
\end{eqnarray}

\noindent are related with the reduced transversal strength and magnetic fields ${j_{h_{-\alpha}}}, {b_{h_{-\alpha}}}$ and with the Rabi frequencies involved ${\Delta_h}_\alpha^\pm$, as they are presented in \cite{delgadoA}. At the same time, these parameters are finally expressed in terms of the physical parameters $t, J_i, B_k, i=1,2,3; k=1,2$. Thus $U(t)=\bigoplus^2_{j=1} {s_{h_{j}}}$, stating a semi-direct product $SU(4) = U(1) \times SU(2)^2$ for $U(t)$ with the Bell states as general basis for all cases $h=1,2,3$. This dynamics mixes the selectable pairs of Bell states in a programmed way. Thus, in this work we explode last property to show that an alternative set of universal gates can be easily constructed. These gates operate on these subspaces defined by the pairs of Bell states as a privileged grammar instead or alternatively to the computational basis (through the block forms depicted there). The second section presents the Boykin et al universal gates and the proposed alternative gates for the current systems. The third section shows the prescriptions to generate the set of universal gates, discussing some possible issues. The fourth section presents a general strategy to evaluate the performance for all gates proposed. Last section states the conclusions and extensions.

\section{Boykin set of universal gates and alternative gates for the Heisenberg-Ising interactions}
It has been shown that two level quantum channel processing  is universal in the quantum gate version of quantum computation \cite{reck1}. In addition, a universal set of two level gates for the computational basis was given by Boykin et al \cite{boykin1}: ${\mathcal B} \equiv \{ S_{\pi/8}, S_{\pi/4}, H, C^a NOT_b \}$ (Table 1). Boykin et al gates are really universal for $U(4)$ operations on quantum information, independently from the basis being used. Then, their form can be used for another arbitrary basis on two levels where quantum computation is being settled. For the Bell states basis, an analog set of gates could be alternative to the last universal set. These gates operates on the entire quantum information space of the bipartite system: ${\mathcal D}=\{ ({\bf 1}_1 \otimes {S_{\pi/8}}_2)_{\rm B} , ({\bf 1}_1 \otimes {S_{\pi/4}}_2)_{\rm B} , ({S_{\pi/8}}_1 \otimes {\bf 1}_2)_{\rm B} , ({S_{\pi/4}}_1 \otimes {\bf 1}_2)_{\rm B}, ({\bf 1}_1 \otimes H_2)_{\rm B} , (H_1 \otimes {\bf 1}_2)_{\rm B} , (C^1 NOT_2)_{\rm B} , (C^2 NOT_1)_{\rm B} \}$. Gate subscript remarks that those forms are written for the Bell basis. Thus, they need be considered covering the two quantum channels simultaneously. For this reason, the graph of any gate in ${\mathcal D}$ comprises always the two channels (Table 2). In particular, the classical symbol for $CNOT$ gate appears horizontally (as a symbolic issue) in the corresponding graphs, due to they are controlled with respect to the quantum information states, not with respect to the single qubit states. 

\begin{center}
\begin{table}[h]
\caption{\label{tab1} Universal set of 1 and 2 level gates for the computational basis.}
%\footnotesize\rm
\centering
\begin{tabular}{l l c c c }
\br
Levels & Gate &  Symbol & Graph & Matrix form \\ 
\br \noalign{\smallskip}
1-Level & $\frac{\pi}{8}$ gate &  $S_{\pi/8}$ &
\begin{minipage}{0.2\linewidth} 
\centering
{\includegraphics[width=16mm, height=8mm]{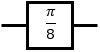}}
\end{minipage}
& ${\tiny { \left(
\begin{array}{cc}
1 & 0   \\
0 & e^{i \pi / 4}    
\end{array}
\right) } \simeq 
{\left(
\begin{array}{cc}
e^{-i \pi/8} & 0   \\
0 & e^{i \pi/8}     
\end{array}
\right) }}$ \\ \noalign{\smallskip}
& $\frac{\pi}{4}$ gate &  $S_{\pi/4}$ &
\begin{minipage}{0.2\linewidth} 
\centering
{\includegraphics[width=16mm, height=8mm]{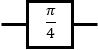}}
\end{minipage}
& ${\tiny { \left(
\begin{array}{cc}
1 & 0   \\
0 & e^{i \pi / 2}    
\end{array}
\right) } \simeq 
{\left(
\begin{array}{cc}
e^{-i \pi/4} & 0   \\
0 & e^{i \pi/4}     
\end{array}
\right) }}$ \\  \noalign{\smallskip}
& {\rm Hadamard} & $H$ & 
\begin{minipage}{0.2\linewidth} 
\centering
{\includegraphics[width=16mm, height=8mm]{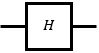}}
\end{minipage}
& ${\tiny \frac{1}{\sqrt{2}} { \left(
\begin{array}{cc}
1 & 1   \\
1 & -1    
\end{array}
\right) }}$ \\ \hline \noalign{\smallskip}
2-Level & {\rm Controlled} & $C^1 NOT_2$ &
\begin{minipage}{0.2\linewidth} 
\centering
{\includegraphics[width=16mm, height=9mm]{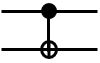}}
\end{minipage}
& ${\tiny { \left(
\begin{array}{cccc}
1 & 0 & 0 & 0   \\
0 & 1 & 0 & 0   \\
0 & 0 & 0 & 1   \\
0 & 0 & 1 & 0   
\end{array}
\right) }}$ \\ \noalign{\smallskip}
& & $C^2 NOT_1$ & 
\begin{minipage}{0.2\linewidth} 
\centering
{\includegraphics[width=16mm, height=9mm]{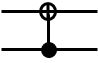}}
\end{minipage}
& ${\tiny { \left(
\begin{array}{cccc}
1 & 0 & 0 & 0   \\
0 & 0 & 0 & 1   \\
0 & 0 & 1 & 0   \\
0 & 1 & 0 & 0   
\end{array}
\right) }}$ \\                    
\br
\end{tabular}
\end{table}
\end{center}

This structure particularly shows how these gates operate on the quantum information being settled in the system and not properly on the physical system. $S_\phi$ gate includes $S_{\pi/8}$ and $S_{\pi/4}$ gates for the corresponding values of $\phi$. In particular note that $(H_1 \otimes {\bf 1}_2)_{\rm B}$ works as a translator between the computational and the Bell basis, closing the equivalence with the ${\mathcal B}$ set. The relevant aspect is the simple construction of these gates for the current Hamiltonian in this work. At the first glance, one can see the correspondence between their block structure of ${\mathcal D}$ elements and the general block in (\ref{sector}). In the next section we state the main concrete prescriptions for each gate.

\begin{center}
\begin{table}[h]
\caption{\label{tab2} Alternative universal set of 2-level gates for the Bell basis.}
%\footnotesize\rm
\centering
\begin{tabular}{l c c c c c c }
\br
Gate & Subspaces & Graph & Matrix form on Bell basis \\
\br
$({\bf 1}_1 \otimes {S_{\phi}}_2)_{\rm B}$ & 1 & 
\begin{minipage}{0.1\linewidth} 
\centering
{\includegraphics[width=16mm, height=8mm]{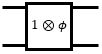}}
\end{minipage} 
& ${\tiny \left(
\begin{array}{cccc}
e^{-i \phi} & 0 & 0 & 0   \\
0 & e^{i \phi} & 0 & 0   \\
0 & 0 & e^{-i \phi} & 0   \\
0 & 0 & 0 & e^{-i \phi}    
\end{array}
\right) }$ \\ \noalign{\smallskip}
$({S_{\phi}}_1 \otimes {\bf 1}_2)_{\rm B}$ & 1 &
\begin{minipage}{0.1\linewidth} 
\centering
{\includegraphics[width=16mm, height=8mm]{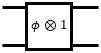}}
\end{minipage}  
& ${\tiny \left(
\begin{array}{cccc}
e^{-i \phi} & 0 & 0 & 0   \\
0 & e^{-i \phi} & 0 & 0   \\
0 & 0 & e^{i \phi} & 0   \\
0 & 0 & 0 & e^{i \phi}    
\end{array}
\right) }$ \\ \noalign{\smallskip}
$({\bf 1}_1 \otimes H_2)_{\rm B}$ & 1 &
\begin{minipage}{0.1\linewidth} 
\centering
{\includegraphics[width=16mm, height=8mm]{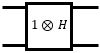}}
\end{minipage} 
& ${\tiny \frac{1}{\sqrt{2}} { \left(
\begin{array}{cccc}
1 & 1 & 0 & 0   \\
1 & -1 & 0 & 0   \\
0 & 0 & 1 & 1   \\
0 & 0 & 1 & -1    
\end{array}
\right) }}$ \\ \noalign{\smallskip}
$(H_1 \otimes {\bf 1}_2)_{\rm B}$ & 1 &
\begin{minipage}{0.1\linewidth} 
\centering
{\includegraphics[width=16mm, height=8mm]{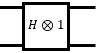}}
\end{minipage} 
& ${\tiny \frac{1}{\sqrt{2}}
{\left(
\begin{array}{cccc}
1 & 0 & 1 & 0   \\
0 & 1 & 0 & 1   \\
1 & 0 & -1 & 0   \\
0 & 1 & 0 & -1  
\end{array}
\right) }}$ \\ \noalign{\smallskip}
$(C^1 NOT_2)_{\rm B}$ & 2 &
\begin{minipage}{0.1\linewidth} 
\centering
{\includegraphics[width=16mm, height=8mm]{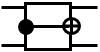}}
\end{minipage} 
& ${\tiny \left(
\begin{array}{cccc}
1 & 0 & 0 & 0   \\
0 & 1 & 0 & 0   \\
0 & 0 & 0 & 1   \\
0 & 0 & 1 & 0    
\end{array}
\right) }$ \\ \noalign{\smallskip}
$(C^2 NOT_1)_{\rm B}$ & 2 &
\begin{minipage}{0.1\linewidth} 
\centering
{\includegraphics[width=16mm, height=8mm]{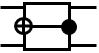}}
\end{minipage} 
& ${\tiny \left(
\begin{array}{cccc}
1 & 0 & 0 & 0   \\
0 & 0 & 0 & 1   \\
0 & 0 & 1 & 0   \\
0 & 1 & 0 & 0  
\end{array}
\right) }$ \\
\br \noalign{\smallskip}
\end{tabular}
\end{table}
\end{center}

The demonstration that ${\mathcal D}$ is universal is trivial, which is based on the universality of ${\mathcal B}$ \cite{boykin1}. Note that due any 2-level gate $G$ can be expressed as a product of other unitary operations in ${\mathcal B}$: $G=\Pi_{i=1}^N g_i, g_i \in {\mathcal B}$. Then, as ${\mathcal T} \equiv (H_1 \otimes {\bf 1}_2)_{\rm B}$ is an operation fulfilling: a) ${\mathcal T}^\dagger={\mathcal T}$, b) ${\mathcal T} \left| \beta_{ij} \right> = \left| i, i \oplus j \right>, {\mathcal T} \left| \beta_{i, i \oplus j} \right> = \left| i, j \right>$. Thus, clearly $G={\mathcal T} (\Pi_{i=1}^N {\mathcal T} g_i {\mathcal T}) {\mathcal T}$ and each ${\mathcal T} g_i {\mathcal T}$ is some gate expressed in Bell basis, a gate operating on two levels and then able to be expressed (at least asymptotically \cite{boykin1}) as a product of elements of ${\mathcal D}$: ${\mathcal T} g_i {\mathcal T}=\Pi_{j=1}^M u^i_j, u^i_j \in {\mathcal D }$. With this, $G={\mathcal T} (\Pi_{i=1}^N \Pi_{j=1}^M u^i_j ) {\mathcal T}$. The process of translation is shown in the Figure 1.

\begin{figure}[th]
\begin{picture}(130,100)(-125,65)
\put(0,0){\makebox(185,55){\vspace*{6cm}
\scalebox{0.9}[0.9]{
\includegraphics[width=13cm]{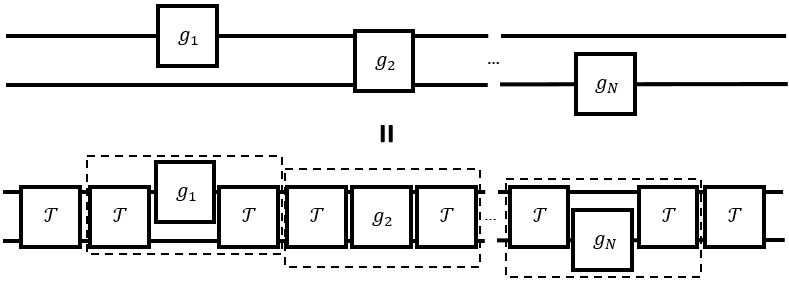}}}}
\end{picture}
\caption{\label{fig1} Quantum gate array showing the equivalence between a traditional circuit using the computational basis and another using the translator gate ${\mathcal T}$. Each block in the dotted boxes can be expressed in terms of elements in ${\mathcal D}$.}
\end{figure}

\section{Heisenberg-Ising gate's realization and possible issues}
All last gates are achievable for the Hamiltonian being considered and its evolution matrix reducible to the blocks (\ref{sector}), through the prescriptions given in the Table 3. This table contains only the immediate prescriptions. 
 
\begin{center}
\begin{table}[h]
\caption{\label{tab3} Basic design parameters for the universal gates in ${\mathcal U}$. There $m, m' \in {\bf Z}$.}
%\footnotesize\rm
\centering
\begin{tabular}{l c c c c c c }
\br
Gate & $({\bf 1}_1 \otimes {S_{\phi}}_2)_{\rm B}$ & $({S_{\phi}}_1 \otimes {\bf 1}_2)_{\rm B}$ & $({\bf 1}_1 \otimes H_2)_{\rm B}$ & $(H_1 \otimes {\bf 1}_2)_{\rm B}$ & $(C^1 NOT_2)_{\rm B}$ & $(C^2 NOT_1)_{\rm B}$ \\ 
\br \noalign{\smallskip}
$h$ & 1 & 1 & 1 & 3 & 1 & 3 \\ \hline \noalign{\smallskip}
${\Delta_h^+}_\alpha$ & $2 \pi$ & $\phi$ & $\pi/2$ & $\pi/2$ & $\pi/4$ & $\pi/4$ \\ \hline \noalign{\smallskip}
${\Delta_h^-}_\alpha$ & $\phi$ & $2 \pi$ & $\pi/2$ & $\pi/2$ & $2m\pi$ & $2m\pi$ \\ \hline \noalign{\smallskip}
${\Delta_h^-}_{-\alpha}$ & $\phi$ & $2 \pi$ & $\pi/2$ & $\pi/2$ & $\pi/2+2m'\pi$ & $\pi/2+2m'\pi$ \\ \hline \noalign{\smallskip}
${b_h}_{-\alpha}$ & 0 & - & $q {\beta j_h}_{-\alpha}$ & $-q \beta {j_h}_{-\alpha}$ & - & - \\ \hline \noalign{\smallskip}
${b_h}_{\alpha}$ & 0 & - & $q {\beta j_h}_\alpha$ & $-q \beta {j_h}_\alpha$ & $|{b_h}_{\alpha}| \rightarrow 1$ & $|{b_h}_{\alpha}| \rightarrow 1$ \\ \hline \noalign{\smallskip}
${j_h}_{-\alpha}$ & $\beta$ & - & - & - & 0 & 0 \\ \hline \noalign{\smallskip}
${j_h}_{\alpha}$ & $\beta$ & - & - & - & 0 & 0 \\ \hline \noalign{\smallskip}
{\rm Other} & - & - & - & - & $m,m' \rightarrow \infty$ & $m,m' \rightarrow \infty$ \\ \noalign{\smallskip}
\br
\end{tabular}
\end{table}
\end{center}

Specific prescriptions should be still written in a more complicated way for physical parameters $t, {B_h}_{\pm \alpha}, {B_h}_{\pm \alpha}$, nevertheless all them are possible and compatible. Some variations could be possible or necessary, but last prescriptions resume the general conditions to reproduce the universal gates set ${\mathcal D}$ on the Bell states. Due to the extension of these prescriptions  they are omitted here. $\alpha$ corresponds always to the first block $j=1$. Note particularly than prescriptions for $({\bf 1}_1 \otimes {S_{\phi}}_2)_{\rm B}$ could be used to generate $({S_{\phi}}_1 \otimes {\bf 1}_2)_{\rm B}$ too when $h=1$ is changed by $h=3$ and $\beta \rightarrow -\beta$ (as for other gates). Nevertheless, we report in the Table 1 an easier implementation for $h=1$. Condition $m \rightarrow \infty$ is optional, but it has been used in the current analysis, indicating the necessity to increase sufficiently the magnetic field on the qubits to approach the evolution matrix asymptotically to $(C^a NOT_b)_{\rm B}$. 

\section{Error estimation strategy based on quantum fidelity}
In order to round the gates proposal, an strategy to evaluate their performance should be delivered. Thus, to set a stability degree for each element of the universal set ${\mathcal D}$, we consider the initial multipartite state (bipartite is sufficient for our purposes):

\begin{eqnarray}
\left| \Psi_0 \right> = \bigoplus_k \sum_j \alpha_{k,j} \left| \psi_{k,j} \right>
\end{eqnarray}

\noindent which is explicitly exhibiting the direct sum structure of the Hilbert space in the current case. Note that $\left| \psi_{k,j} \right>$ is the basis being used in the representation (some arrangement of Bell states, dependent on $h$ in our case), with each element labeled by the block $k$ and a label  $j$ for each element of the pair. Under the effect of some gate on $SU(4): U = \bigoplus_k s_{h_k} \in {\mathcal D}$, where $s_{h_k}$ is the $U(2)=U(1) \times SU(2)$ block (\ref{sector}). Here, $h$ is an additional label identifying different decompositions and arrangements of $\left| \psi_{k,j} \right>$ (the direction in which the control field is applied, in our case) and $k$ is the block number. Then, the transformed state $\left| \Psi_f \right>$ under the gate $U$ is:

\begin{eqnarray}
\left| \Psi_f \right> = U \left| \Psi_0 \right> = \bigoplus_k \sum_j \alpha_{k,j} s_{h_k} \left| \psi_{k,j} \right>
\end{eqnarray}

Now, we are interested in a tiny failure in the gate prescriptions, generating an alternative effect but very near from the proposed gate,
$U' = \bigoplus_k s'_{h_k} = \bigoplus_k s_{h_k} + \delta s_{h_k}$, where $\delta s_{h_k}$ is a tiny variation of $s_{h_k}$ generated by the variation of some of their parameter prescriptions ${\bf p}=(p_1,p_2,...,p_N)$ (${\bf p}=(t,B_{1_h},B_{2_h},J_1,J_2,J_3)$ in our case). This gate generates on $\left| \Psi_0 \right>$:

\begin{eqnarray}
\left| \Psi'_f \right> = U' \left| \Psi_0 \right> = \left| \Psi_f \right> + \bigoplus_k \sum_j \alpha_{k,j} \delta s_{h_k} \left| \psi_{k,j} \right>
\end{eqnarray} 

\noindent Gate stability can be quantified by its fidelity \cite{delgadoB}. Developing $\delta s_{h_k}$ to second order:

\begin{eqnarray}\label{derivada}
\delta s_{h_k} &\approx& {\mathcal D} s_{h_k} + \frac{1}{2}  {\mathcal D}^2 s_{h_k}, \quad {\mathcal D} = d{\bf p} \cdot \nabla_{\bf p}  \nonumber
\end{eqnarray} 

\noindent  and if $\rho_f = \left| \Psi_f \right>\left< \Psi_f \right|, \rho'_f = \left| \Psi'_f \right>\left< \Psi'_f \right|$, then:

\begin{eqnarray}
{\mathcal F}^2 = {\rm Tr} (\rho_f \rho'_f) = |1+\sum_{k,j,j'} \alpha^*_{k,j'} \alpha_{k,j'} \left< \psi_{k,j'} | s_{h_k}^\dagger \delta s_{h_k} | \psi_{k,j} \right>|^2
\end{eqnarray} 

Developing the last expression using (\ref{derivada}) and considering $s_{h_k} s_{h_k}^{\dagger} = {\bf 1}_k$, it gives:

\begin{eqnarray}\label{f2}
{\mathcal F}^2 &=& 1-\sum_{k,j,j'} \alpha^*_{k,j'} \alpha_{k,j'} a_{k,j,j'} + |\sum_{k,j,j'} \alpha^*_{k,j'} \alpha_{k,j'} b_{k,j,j'}|^2 \\
&{\rm with:}& a_{k,j,j'} = \left< \psi'_{k,j} | {\mathcal D} s_{h_k}^{\dagger} {\mathcal D} s_{h_k} |\psi'_{k,j} \right>, b_{k,j,j'} = \left< \psi'_{k,j} | s_{h_k}^{\dagger} {\mathcal D} s_{h_k} |\psi'_{k,j} \right> \nonumber
\end{eqnarray} 

\noindent outstandingly, ${\mathcal F}^2$ depends quadratically from $d {\bf p}$, a property due to ${s_h}_j$ is unitary. This expression set a procedure to evaluate the performance of any element in ${\mathcal D}$ for any state $\left| \Psi_0 \right>$ being processed around of their specific prescriptions given in the Table 3. This task implies evaluate ${\mathcal F}^2$ for each gate in ${\mathcal D}$ and any state $\left| \Psi_0 \right>$ in the Hilbert space ${\mathcal H}^2$.

\section{Conclusions}

Exact control for time independent magnetic fields has been used here to reduce the evolution blocks to pertinent blocks to reproduce the gates ${\mathcal D}$ alternative to ${\mathcal B}$, but other kinds of control (optimal, non-resonant, etc.) could be introduced \cite{dales2,boscain1}, whose control schemes are well known for the $SU(2)$ dynamics. More achievable forms for magnetic fields pulses are in order to reproduce the proposed gates, particularly avoiding resonance effects.

Nevertheless gate fidelity has been theoretically quantified, additional concrete research is necessary to analyze this fidelity for the current constructions when the prescriptions are slightly modified due to the main uncontrollable factors. In particular, as time and strength of magnetic fields are currently well controlled \cite{sekat1}, possibly the most sensible factors in the model are the strengths $J_i$ for the non-local interactions in the Heisenberg-Ising model \cite{delgado0}. A complete analysis based on the formula (\ref{f2}) for each gate is currently in progress. 

This proposal is based on Bell states instead of single qubits states, which are technologically a challenge in terms of experimental stability and coherence. Despite, this states appears as better candidates to set a computational grammar on magnetic systems easing quantum control and quantum gate engineering. In addition, the approach is clearly settled on the quantum information manipulation more than in the physical systems where it lays. 

\section*{Acknowledgements}
The support from Tecnol\'ogico de Monterrey to develop this research is acknowledged.

\end{document}